\documentclass[prl, preprint, floatfix]{revtex4}
\usepackage{amsmath, amssymb}
\usepackage{graphicx}
\usepackage{natbib}

\newcommand{\ba}{\begin{eqnarray}}
\newcommand{\ea}{\end{eqnarray}}

\newcommand{\be}{\begin{equation}}
\newcommand{\ee}{\end{equation}}

\begin{document}
\hyphenation{e-lec-tro-wet-ting}
\title{Microscopic structure of electrowetting-driven transitions on superhydrophobic surfaces}
\author{A. Staicu}
\email{a.d.staicu@tnw.utwente.nl}
\author{G. Manukyan}
\author{F. Mugele}
\affiliation{Physics of Complex Fluids, Faculty of Science and Technology, University of Twente, PO Box 217, 7500AE Enschede, The Netherlands}
\date{\today}
%
%abstract..................................................
\begin{abstract}
We investigate directly at the microscale the morphology of the
electrowetting induced transition between the Cassie-Baxter and Wenzel
states for a water droplet on a superhydrophobic surface. Our experiments demonstrate that the transition originates in a very narrow annular region near the macroscopic contact line, which
is first invaded by water and causes a thin film of air to be
entrapped below. At high applied voltages, a growing fraction of
microscopic air-pockets collapse, resulting in a partial Wenzel
state. Modulations in the intensity of the light reflected from
individual micro-menisci clarify that the local contact angles near
the filling transition are close to the usual advancing values for
contact lines on smooth surfaces.
\end{abstract}
%PACS numbers? search the ones for superhydrophobic
%\pacs{68.15.+e, 68.45.Gd, 83.60.Np}
\maketitle
It is well known that the presence of roughness at microscale, whether
naturally occurring or artificially created, leads to the phenomenon of
superhydrophobicity\,\cite{quere1}. Depending on whether or not the liquid
fills the crevasses of the microscopic texture (typically
microscopic pillars uniformly spaced), the mobility of water droplets
on such surfaces can be either severely decreased, resulting in
sticking drops (Wenzel state\cite{wenzel}) or, respectively, greatly
enhanced (Cassie-Baxter state\cite{cassie}).
% The essential required ingredients are:
% effective control of interfacial energies (e.g. tunable liquid-solid
% interfacial energy) and understanding of the parameter space that
% determined the system being in either of the two metastable states.
%..................................................
\begin{figure}[b]
\begin{center}
	\includegraphics[height = 4cm]{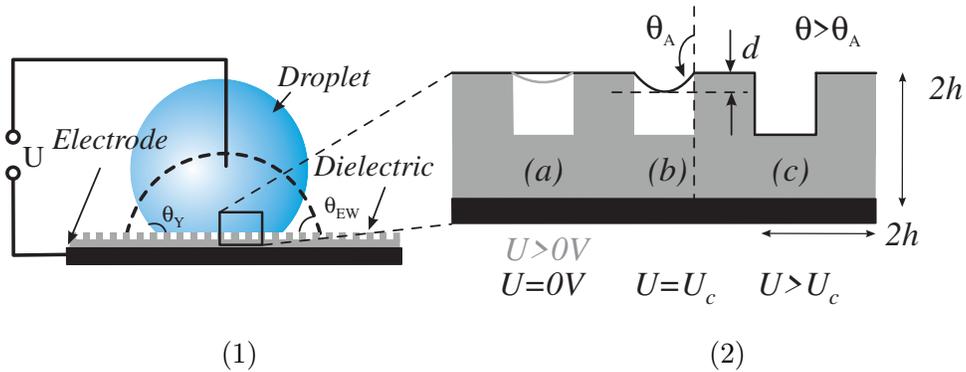}\\
(1)\hspace{6cm}(2)
\end{center}
\caption{(1) Electrowetting: a voltage $U$ is applied between a droplet and an electrode covered with a dielectric micro-patterned surface. The contact area is viewed with an inverted microscope through the superhydrophobic substrate . (2) Scenario for the filling with water of individual micro-pits: (a) $U=0$ water menisci are flat and become increasingly curved for $0<U<U_c$ (critical voltage); (b) for $U_c$ such that the contact angle at the vertical walls is the advancing angle $\approx \theta_A$ the filling process starts; (c) for $U>U_c$ the menisci advance to the Wenzel state, provided the opposing hydrostatic pressure does not change.}
\label{fig_exp}
\end{figure}
%..................................................
% control of wetting
Effective control of water wettability on smooth surfaces has been
demonstrated in material research already for a long time (by external
action on droplets by electric fields, light, temperature and
chemistry), but only recently these effects have been combined with
textured superhydrophobic surfaces in order to drive the transition
between the Cassie-Baxter, or air-pocket state, and Wenzel state, when
liquid fills the space between the micro-pillars
(Fig.\ref{fig_exp}). A situation of considerable interest for
industrial applications would be to be able to quickly and reversibly
switch between these wetting states in a non-intrusive manner. Recent
promising efforts focus on the use of the electrowetting effect (EW).
Several studies\,\cite{Krupenkin2004, herbertson, dhindsa, verplanck,
wang, zhu}) reported EW induced transitions from highly mobile
droplets to pinned states on microstructured surfaces. These were
signaled by more or less abrupt jumps in the observed value of the
macroscopic contact angle, which in the absence of other indicators is
qualitatively associated with the Wenzel state.  Typically, when air
is the ambient fluid, the transition is irreversible, although
external intervention\,\cite{Krupenkin2004} can return the system back
to the Cassie state. Irreversibility is attributed mainly to the lower
energy of the Wenzel configuration and/or contact angle
hysteresis. When experiments are performed in an oil
environment\cite{dhindsa}, the reversibility is achieved simply by
removing the applied voltage.It has been been speculated that it may
be due to the presence of residual oil under the droplet while in
Wenzel state. A common aspect of previous EW studies is that they only
provide a {\em macroscopic} description of the transition.  We believe
that access to the microscale is the key to solving problems such as
the transition threshold, the dynamics of the droplet-surface liquid
invasion, and ultimately, physical understanding of the
irreversibility puzzle.  Using this approach outside of the
electrowetting context, a recent study \cite{mauro} showed that, when
the wetting transition is spontaneous, the Wenzel state (lower-energy)
is first nucleated at a single interior point from which a wetted area
expands radially. The viscous dissipation and roughness are important
for determining the shape of the Wenzel wetting fronts. It is unclear
if this picture is valid in the presence of electric fields.

%..................................................
\begin{figure}[t]
\begin{minipage}[t]{.95\linewidth}
\begin{center}
\includegraphics[width = 13cm, keepaspectratio=true, clip]{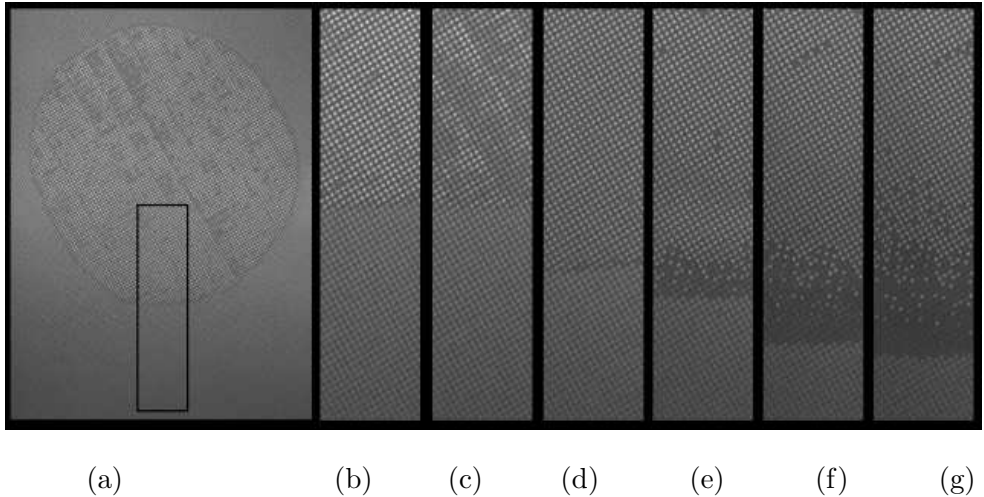}
\end{center}
\end{minipage}
\begin{minipage}{.98\linewidth}
\hspace{1cm}(a)\hspace{2.8cm}(b)\hspace{1cm}(c)\hspace{0.9cm} (d)\hspace{1.2cm}(e)\hspace{1.2cm}(f)\hspace{1.2cm}(g)
\end{minipage}
%..................................................
\caption{View of the droplet-superhydrophobic substrate contact area. Voltage increases from left to right : (a) 0V - Cassie state, the entire contact area is shown: the droplet touches the tops of the microscopic pillars and rests on a cushion of air, (b-g) Magnified view of the  black rectangle region in (a) for voltages $U = 0...250$ ($50$V steps), taken during a $10$ second voltage ramp. Around $U_c\approx 150$V  the first micro-pits go to the Wenzel state (see also Fig.\protect\ref{fig_wetting}).} 
% $U_c$ - The area near the apparent
% circular contact line is ``impregnated'', (c) $U>U_c$ the rest of the
% droplet-substrae contact is abruptly invaded: the final state is a
% mixed Cassie-Wenzel state. }
%
\label{fig_transition}
\end{figure}
%..................................................
%
In this paper we investigate experimentally the morphology of the
Cassie-Wenzel EW-induced transition by focusing on the
droplet-substrate contact area by using reflection
microscopy. In this manner, we can infer the shape of the microscopic water-air menisci formed between the microstructures during the EW-transition. To control the transition itself, the voltage applied between a drop of conductive liquid and a insulator-covered superhydrophobic
surface is gradually increased, leading to an increased affinity between
the drop and the substrate.

The experimental setup (Fig.\,\ref{fig_exp}) follows the classic EW
configuration\,\cite{mugele_rev}: a millimeter-sized water drop sits
on a ITO covered glass wafer on which a dielectric layer is applied,
with the difference that here the dielectric has a micro-patterned
structure.  To obtain it, the wafer is first spin-coated with a
SU-8\footnote{standard polymeric photo-resist used in
micro-lithography} layer of uniform thickness $h=5\mu$m. Using
micro-lithography a periodic patternof cylindrical micro-posts
arranged in a square pattern (Fig.\,\ref{fig_exp}) is developed from
an additional $h=5\mu$m spin-coated SU-8 layer. The solid fraction $\phi$ of
this geometry is $\phi=0.196$ and the roughness (total pattern
area/normal projection area) is $r=1.785$. The entire structure is
made hydrophobic by slowly dip-coating from a $6$\verb+%+ Teflon AF
(Dupont) solution diluted {$10\times$} in FC75. Electron microscope
pictures (Fig.\,\ref{fig_exp}b) show that this coating is does not
significantly change the original geometry of the microscopic pillars.
The use of this construction method ensures that the substrates remain transparent, such that we
can observe the water droplet-substrate contact region from below using an
inverted microscope in reflection mode. The sample was illuminated
with collimated green light $\lambda =520$nm and the system is in ambient air. A platinum electrode connects the water drop to a AC voltage
source ($U_{max}$ 300V, frequency $f=5\dots 10$kHz) and the electrical circuit is closed through the conducting ITO
substrate.

On flat SU-8 surfaces, the Teflon AF coating provides a Young
angle of $\theta_{Yair} \approx 120^\circ$. For a micro-patterned
substrate with a fraction of $\phi$ of the surface in liquid-solid
contact, the macroscopic contact angle of the Cassie-Baxter state is
$\cos\theta_{CB}= \phi\cdot\cos\theta_Y-1+\phi$
which is in agreement with our measurements (for $\phi=0.196$ the measured angle
is $\theta^m_{CB}\approx 144^\circ$, while the theoretical one is
$\theta^t_{CB}\approx 149^\circ$ ).
%
% As the lateral features of EW-enabled superhydrophobic surfaces are
% always comparable to the thickness of the dielectric layer, this makes
% the morphology of the transition even more difficult to anticipate.
When an AC voltage (5khz) is applied and slowly increased, the
droplet-substrate circular contact area is observed to increase
(droplet spreads).The intensity of the light reflected from
individual air-liquid interfaces between the micro-pillars passes
through minima and maxima, indicating that air is gradually being
replaced from below the droplet. As the voltage reaches a
critical value $U>U_c\approx 150$V, we observe at the {\em periphery}
of the droplet-substrate contact area (Figs.\ref{fig_transition}(d)
and \ref{fig_wetting}) that water invades a single ring of microscopic
pits. This ring automatically forces a volume of ambient fluid to be
entrapped below. This discrepancy with the
random-point nucleation observed in spontaneous wetting \cite{mauro} is explained by the fact that at the droplet apparent contact line the electric fields are enhanced. A quick
estimate follows from the Appendix of \cite{Vallet1999}: $E=E_a+U/\sqrt{\pi h_{e} \rho}$,
where the curvature $\rho\sim h_{e}$. Here $h_e$ is the effective thickness of the equivalent capacitor below the drop, {\em i.e.} the edge electric fields are $\approx \sqrt{\pi}$ higher than the average interior ones. Above $U_c$ the droplet
continues to spread and the newly covered surface is in the
Wenzel state; additionally, interior micro-pits are also invaded.The trapped air becomes increasingly compressed and makes the process inefficient.
(see Fig.\ref{fig_transition}(e-g)). The entrapped ambient responds dynamically
for experiments with fast voltage ramps ($<<1$
second) or when an ambient fluid with larger
viscosity ({\em e.g.} silicon oil) is used. Then we
observed partial wetting patterns  (Fig.\ref{fig_wetting}) that point to a dynamic
instability, probably similar in origin to those observed on smooth substrates\,\cite{staicu2006}. 
Note that the transition observed here for slow voltage ramps ($T >1$ second) is
``smooth'' : we do not observe a clear jump in the macroscopic contact
angle (Fig.\ref{fig_transition}(d-g) and droplet volume is conserved).
This is not surprising given the one-by-one invasion process we
observe. 
% We will discuss elsewhere that such discontinuous transitions
% are related to the poor quality of the additional hydrophobic layer
% applied on top of the bare micro-structure.

% An alternative path to a complete Wenzel state would be to let the gas in the air-pockets slowly diffuse in the liquid, but for the timescale of our experiments (max 50sec) we have not observed this process. 

\begin{figure}[t]
	\includegraphics[height = 6cm, keepaspectratio=true]{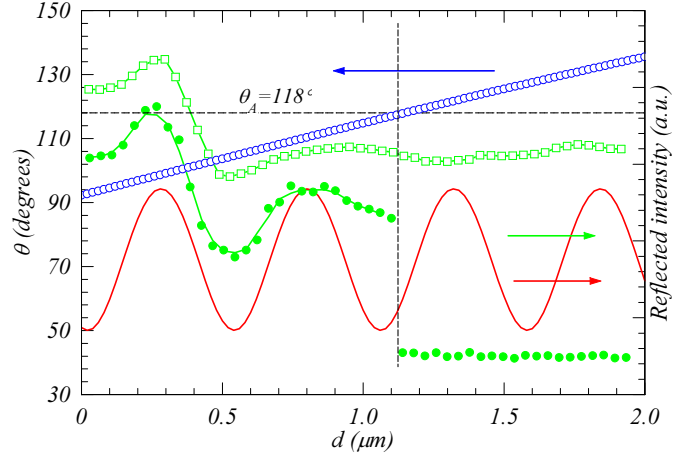}
\caption{Measured light intensity reflected from two selected micro-pits ({\large $\bullet$}) as a function of voltage $U$ and comparison with the calculated values for an air pocket of uniformly decreasing thickness (solid line). The sudden drop in intensity corresponds indicates the abrupt filling (Wenzel transition). For a neighboring pit ({\large $\bullet$} shifted vertically), the entrapped air prevents the transition. The transition is observed when the deflection $d$ from a flat micro-meniscus is $d\approx 1.1\mu m$, which translates to a contact angle $\theta\approx 118^\circ$ ($\circ$ curve).}
\label{fig_fringes}
\end{figure}

For a quantitative relation between $U_c$ and the individual
micro-menisci shapes, we recorded the variation of the reflected light
intensity coming from individual microscopic pits during the
application of a linear voltage ramp to $U>U_c$
(Fig.\ref{fig_fringes}. The interference curves are a measure of the (average) deviation
$d$ of the micro-menisci from flat interfaces (resting atop the
pillars) under the effect of local Maxwell stresses. There are two
possibilities: (1) the menisci are curved downward
(Fig.\ref{fig_exp}(3a, 3b) but don't advance to invade the pits unless
the voltage exceeds $U_c$ and (2) the menisci stay relatively flat and
gradually fill the pits for $U>0$, such that when $U=U_c$ the pits
are completely filled (Fig.\ref{fig_exp}(3c)). This second scenario is eliminated by the small number of
inteference peaks seen before the transition (indicates $d<<h$ at the transition) and reversibility, {\em i.e.} when a voltage below $U_c$ is removed, the light intensity
returns to approximately the $U=0$ value).

%..................................................
\begin{figure}[t]
	\includegraphics[height = 3cm, keepaspectratio=true]{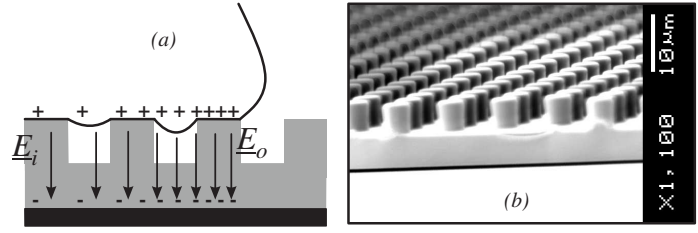}
\caption{(a) Micromenisci situated close to the apparent droplet contact line have higher curvature than the interior ones, as a consequence of the edge electric field enhancement: $E_i > E_o$.}
\label{fig_curv}
\end{figure}
%..................................................

%..................................................
%..................................................
%
%  as indicated in Fig.\ref{fig_reverse}. Here we show the intensity of the light reflected from the microstructures during a triangular voltage ramps with a period of 20 seconds. The intensities recorded at the beginning and end of the ramps are essentially the same, apart from a small hysteresis most likely due to non-flat pillar tips and seen only during the first period. This result clarifies that, technically speaking, the Cassie-Baxter state is maintained up to $U_c$, and, as the voltage increases, the air-water micro-menisci become increasingly bent towards the micro-pits without a significant displacement of the microscopic contact lines. 

We provide here a simple model to estimate the local contact angle at
the mico-pillars for which the transition occurs. For simplicity
we consider that the electric fields $E_a$ are uniform inside the air
pockets and the corresponding Maxwell stresses\cite{buehrle2007} cause
a uniform curvature of the micro-menisci. The two pressures are
related by the Laplace relation: $ p_{el}=\frac{\epsilon_0
E_a^2}{2}\approx\frac{\gamma}{r}$.  For a meniscus radius of curvature
$r$, the local contact angle is given by $\sin(\theta)= - d/(2r)$ and
the maximum deviation from pillar tips is $d =
r(1-\sin(\pi-\theta))$. We will approximate the height of the air
pockets at a given voltage $U$ to be simply $h-d$, where $h$ is the
depth of the micro-pits (Fig.\ref{fig_exp}(3b)).

For a uniformly decreasing height of the air-pockets starting from
$h$, the calculated change in the reflected line intensity is shown in
Fig.\ref{fig_fringes}(b). As we are only interested in the behavior of
$d$ with $U$ and not in the exact functional dependence. This is then compared to two curves from Fig\ref{fig_fringes}(b) from the interior region, one that will fill (filled symbol) and one stable (open symbol). We rescaled the positions of the two measured
interference maxima so that they coincide with the calculated
ones. Then we can simply ``read out'' the $h-d$ value for the transition
and translate it into a curvature using the arguments given
earlier. The contact angle for the transition, corresponding to the
sudden drop in the measured intensity is then $\theta_m \approx
118^\circ$, which is of the order of the advancing contact angle of a
triple water-air-Teflon AF interface on a smooth substrate. Why is the rest of the interior air pockets not filling simultaneously? As some of the air
pockets are filled, the droplet contact line is ``sealed'' and we
expect the pressure in the neighboring ones to increase slightly and
delay the filling of other pits to higher applied voltages. The larger
the water-impregnated area, the more stable the remaining air-pockets
become, which should be the reason for not observing the
full-transition at the highest voltages we applied ($>300$V). The
other possibility is, obviously, the saturation
effect\cite{mugele_rev}. We have noticed that for improperly coated
micro-patterned substrates the transition occurs at a significantly
smaller voltage premature and the inferred critical contact angle for
the transition is close to that of the bare SU-8 surface ($\approx
94^\circ$).

In spite of the robustness of the dielectric, the electrowetting
effect saturated before we could squeeze a significant fraction
of the interior micro-pits to fill up; we have observed that for the
highest voltages, as the trapped air is confined to smaller regions,
it forms shallow bubbles which may rise above the pillar
structure. When the voltage is applied very fast (the extreme case
corresponds to a voltage step), we see organized patches of similar
sizes (Fig.\ref{fig_wetting}) of impregnated micro-pits rather than
isolated single ones (``quasi-static'' case). Additional
experiments performed in silicone oil (Fluka AS4 with viscosity
$\mu \approx 6$mPa$\cdot$s) produced similar results for slower voltage ramps.

%..................................................
\begin{figure}[t]
\centering
\includegraphics[height = 4cm, keepaspectratio=true, clip]{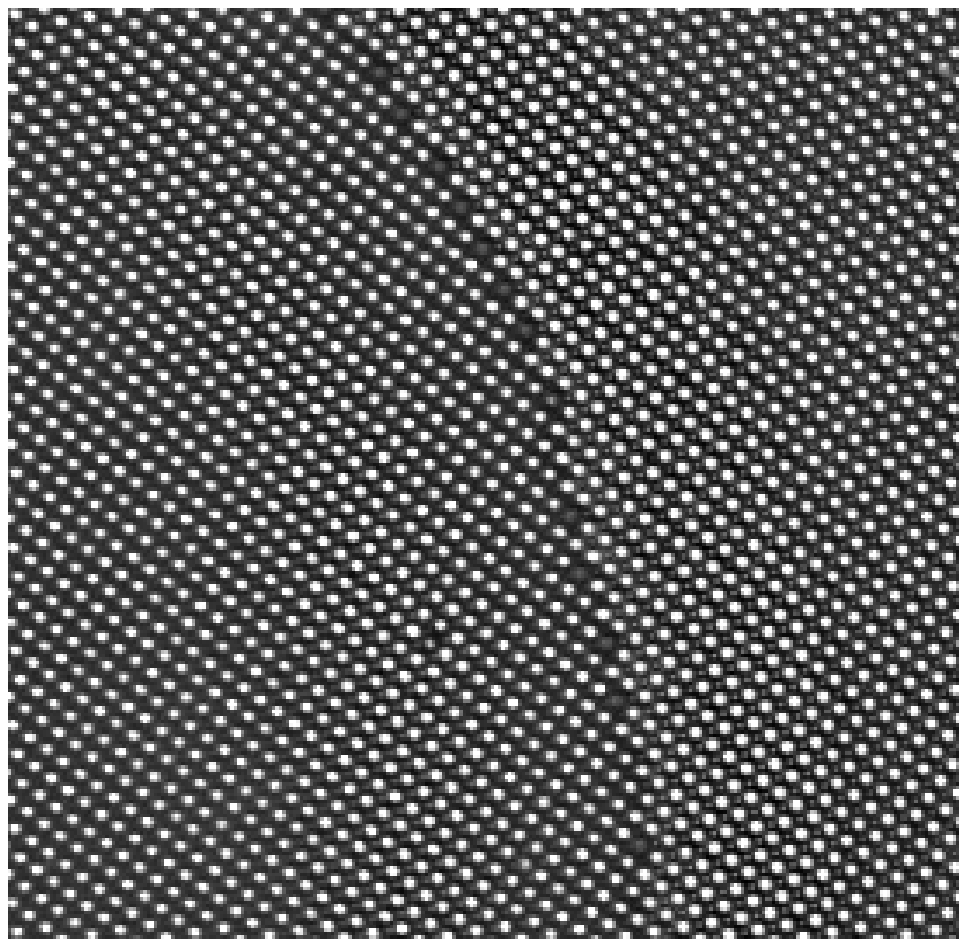}
\includegraphics[height = 4cm, keepaspectratio=true, clip]{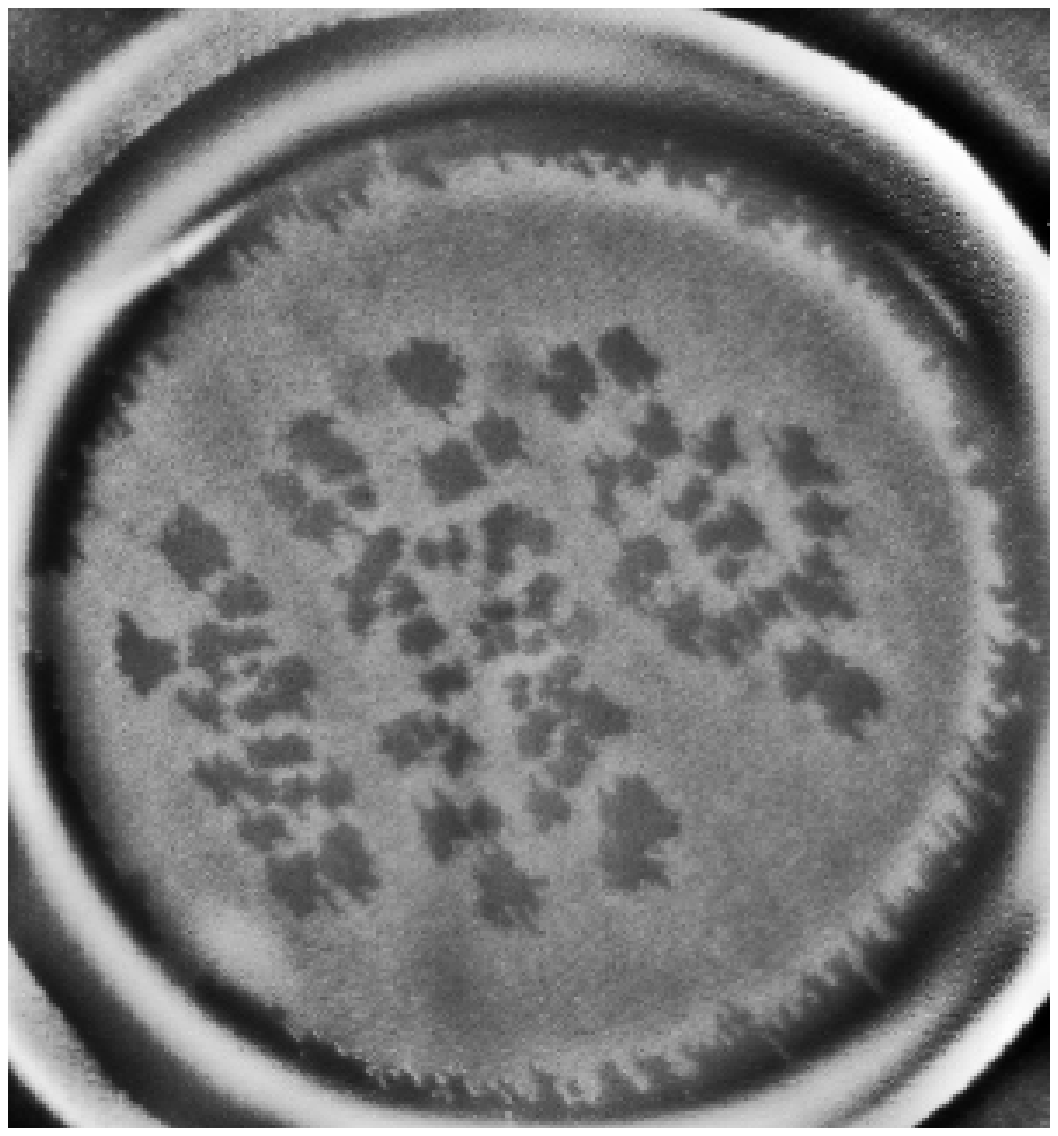}
\includegraphics[height = 4cm, keepaspectratio=true, clip]{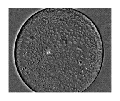}\\
\hspace{4.5cm}(a)\hspace{\stretch{1}}(b)\hspace{\stretch{1}}(c)\hspace{4.5cm}
\caption{(a) Incipient state of the EW-induced wetting transition: due
to increased electric fields at the contact line
(\protect{\ref{fig_curv}}), a single ring of air-pockets collapses
first. (b) Partial Wenzel state wetting pattern obtained after the
application of fast voltage ramps. The dynamics of the process is
essentially the same, but slowed down considerably, when the
experiment is performed in silicon oil environment (c). Imaging
oil-water interfaces is however more difficult due to smaller
refractive index contrast.}
\label{fig_wetting}
\end{figure}
%..................................................
%
To summarize, we have performed the first microscale study of the
morphology and dynamics of the EW-induced wetting transition of a
droplet on a superhydrophobic surface. We have identified that the
transition always originates at the apparent droplet-patterned
substrate contact line, due to locally higher fringe electric fields. The impregnation of the micro-structure with liquid is only partial due to the entrapping of a finite volume of air below the electrowetting drop. This situation is aggravated at high rates of
increasing voltage, when dynamic wetting patterns are
observed. Their formation can be slowed down if an oil is used instead
of air as ambient liquid. In order for a micro-pit to be filled with
water, the curvature caused by the local maxwell stresses has to be
sufficiently high such that the local contact angles exceed the
advancing value in the absence of electric fields. We believe that
this is essentially a nucleated thin-film dewetting of ambient fluid,
initially trapped between the droplet and the substrate.

Financial support within the joint
program Micro- and Nanofluidics by the MESA+ institute for
nanotechnology and by the IMPACT institute at Twente University is
gratefully acknowledged.
%references........................................
\bibliography{biblio_wetting}
\end{document}